\documentclass[aps, prl, twocolumn, showpacs, amssymb, amsmath, aps, superscriptaddress] {revtex4-2}
      
\usepackage[noend]{algpseudocode} 
\usepackage{algorithm} 
\usepackage{amsfonts}
\usepackage{amsmath}
\usepackage{amssymb}
\usepackage[toc,page]{appendix}  
\usepackage{bbold}
\usepackage{bm}  			
\usepackage{color}
\usepackage{dcolumn}  		
\usepackage{epstopdf}
\usepackage[mathscr]{eucal}
\usepackage[hidelinks]{hyperref}
\usepackage{graphicx}  		
\usepackage[utf8]{inputenc}
\usepackage{latexsym}
\usepackage{mathrsfs}
\usepackage{mathtools}
\usepackage{physics}
\usepackage{silence}
\usepackage{soul}
\usepackage{tikz}
\usetikzlibrary{matrix,arrows,decorations.pathmorphing}

\makeatletter 
\def\maketag@@@#1{\hbox{\m@th\normalfont\normalsize#1}} 
\makeatother 

\makeatletter
\newcommand*{\balancecolsandclearpage}{%
  \close@column@grid
  \clearpage
  \twocolumngrid
}
\makeatother

\renewcommand{\Tr}{\mathrm{Tr}}

\begin{document}

\title{Trotter Errors and the Emergence of Chaos in Quantum Simulation}

\author{Kevin W. Kuper}\affiliation{Wyant College of Optical Sciences, University of Arizona, Tucson, AZ 85721, USA}
\author{Jon P. Pajaud}\affiliation{Wyant College of Optical Sciences, University of Arizona, Tucson, AZ 85721, USA}
\author{Karthik Chinni}\affiliation{Center for Quantum Information and Control, Department of Physics and Astronomy, University of New Mexico, Albuquerque, NM 87131, USA}
\author{Pablo M. Poggi}\affiliation{Center for Quantum Information and Control, Department of Physics and Astronomy, University of New Mexico, Albuquerque, NM 87131, USA}
\author{Poul S. Jessen}\affiliation{Wyant College of Optical Sciences, University of Arizona, Tucson, AZ 85721, USA}
\date{\today}

\begin{abstract}
\noindent As noisy intermediate-scale quantum (NISQ) processors increase in size and complexity, their use as general purpose quantum simulators will rely on algorithms based on the Trotter-Suzuki expansion. We run quantum simulations on a small, highly accurate quantum processor, and show how one can optimize simulation accuracy by balancing algorithmic (Trotter) errors against native errors specific to the quantum hardware at hand. We further study the interplay between native errors, Trotter errors, and the emergence of chaos as seen in measurements of a time averaged fidelity-out-of-time-ordered-correlator (FOTOC).
\end{abstract}

\maketitle

\noindent 	
Steady progress in the performance of quantum hardware has led to the point where quantum devices with several tens of qubits \cite{Arute2019, Wu2021} can operate in regimes that challenge verification by classical computation \cite{Feng2022}. This emerging class of Noisy, Intermediate-Scale Quantum (NISQ) processors \cite{Preskill2018} will likely continue to grow in size and complexity, while still falling short of the thresholds for error correction and fault tolerance for some time to come. As such, they may reach new milestones in the quest for quantum supremacy, while also reminding us that the quantum advantage offered by such devices will depend on factors beyond the raw number of qubits \cite{Baldwin2022}. Among these are gate fidelity, circuit connectivity, noise suppression, decoherence, and the computational task itself. 

One of the most promising applications of NISQ hardware is in non-error corrected quantum simulation. The term ``quantum simulator'' suggests a programmable device capable of replicating the dynamics driven by an arbitrary Hamiltonian. In most cases such simulators will operate in the circuit paradigm, with access to a limited set of native Hamiltonians and the gates they generate. At the basic level, the goal is then to obtain a coarse-grained version of the evolution driven by some non-native Hamiltonian $H$ during a time interval from $t=0$ to $T$. One widely used method is to approximate a unitary time step with a Trotter-Suzuki expansion \cite{Lloyd1996, Childs2021}. In the simplest case, if $H$ is the sum of two non-commuting native Hamiltonians $H_1$ and $H_2$ , then $e^{-iH\tau}\approx e^{-iH_1\tau}e^{-iH_2\tau}$ , where $e^{-iH_1\tau}$  and $e^{-iH_2\tau}$  are part of the native gate set.

Trotter errors generally increase with the step size $\tau=T/n$, while native errors grow with the number of gate operations and thus the number of steps $n$ needed for a simulation of length $T$. If so, one expects an optimal step size where the fidelity of the simulation is maximized. However, recent work has emphasized that Trotterization can lead to instability and chaos in the simulation of otherwise integrable Hamiltonians, in which case estimates for the Trotter error are unreliable or too loose \cite{Sieberer2019, Heyl2019, Kargi2021}. In this scenario simulation errors may remain bounded and controlled below a critical step size and increase sharply above it. Therefore, optimizing a Trotterized simulation will involve measuring the level of native errors, balancing these with models of the Trotter errors, and checking if chaos is present at the chosen step size. 

In this letter we explore the interplay between native errors, Trotter errors, the emergence of chaos, and the use of an Out-of-Time-Ordered-Correlator (OTOC) to detect its presence in an experimental setting.  Specifically, we run Trotterized simulations of the well-known Lipkin-Meshkov-Glick (LMG) model on a Small, Highly Accurate Quantum processor, a unique device based on individual Cs atoms in their electronic ground state. While the SHAQ processor does not operate in the circuit paradigm, it is universally programmable via Optimal Control, and offers state-of-the-art control in a $4$-qubit equivalent Hilbert space with dimension $d=16$.  Within this Hilbert space we can prepare any quantum state, implement any unitary map, perform any orthogonal measurement, and carry out quantum simulations with up to $150$ iterations of any chosen unitary map with a fidelity $>0.99$ per step. This makes it an excellent testbed for the study of complex dynamics in quantum systems. Underlying details of its operation are not pertinent here but can be found in the literature \cite{Smith2013, Anderson2015, Lysne2020}.

As a starting point for experiments, we pick the Lipkin-Meshkov-Glick (LMG) model \cite{Lipkin1965, Zibold2010} as the target for simulation. The LMG Hamiltonian describes a spin system and has the form
\begin{align}
\frac{H_{\text{LMG}}}{\gamma}=-(1-s)J_z - \frac{s}{2J} J_x^2,
\end{align}

\noindent were $s$ controls the relative strengths of the linear and nonlinear rotations and $\gamma$ sets the time scale for the evolution. For the remainder of this letter, we set $\hbar=\gamma=1$, making the spin, energy, and simulation time dimensionless. The LMG model is equivalent to a fully connected transverse Ising model and undergoes a ground state quantum phase transition at $s=0.5$ where a single well bifurcates in the classical phase space. The total spin is conserved because $[H_{\rm LMG},J^2]=0$, allowing us to simulate spins up to $J=15/2$ on our processor.  This simple model is convenient for our purposes, as decomposing the evolution driven by this Hamiltonian in a first-order Trotter-Suzuki expansion yields the quantum Kicked Top (QKT), which is a popular paradigm for quantum chaos \cite{Haake1987, Chaudhury2009}. The exact LMG and Trotterized time steps are given by

\begin{align}
&U_{\rm LMG}(\tau)=e^{i[(1-s)J_{\rm z}+(s/2J) J_{\rm x}^2]}\tag{2a},\\
&U_{\rm Trot}(\tau)=e^{i(1-s){\tau}J_{\rm z}}e^{i(s{\tau}/2J) J_{\rm x}^2},\tag{2b}
\end{align}

\noindent where (2b) is easily recognized as a single iteration of a QKT with a non-linear rotation $\propto \tau$ .

\begin{figure}
[t]\resizebox{8.5cm}{!}
{\includegraphics{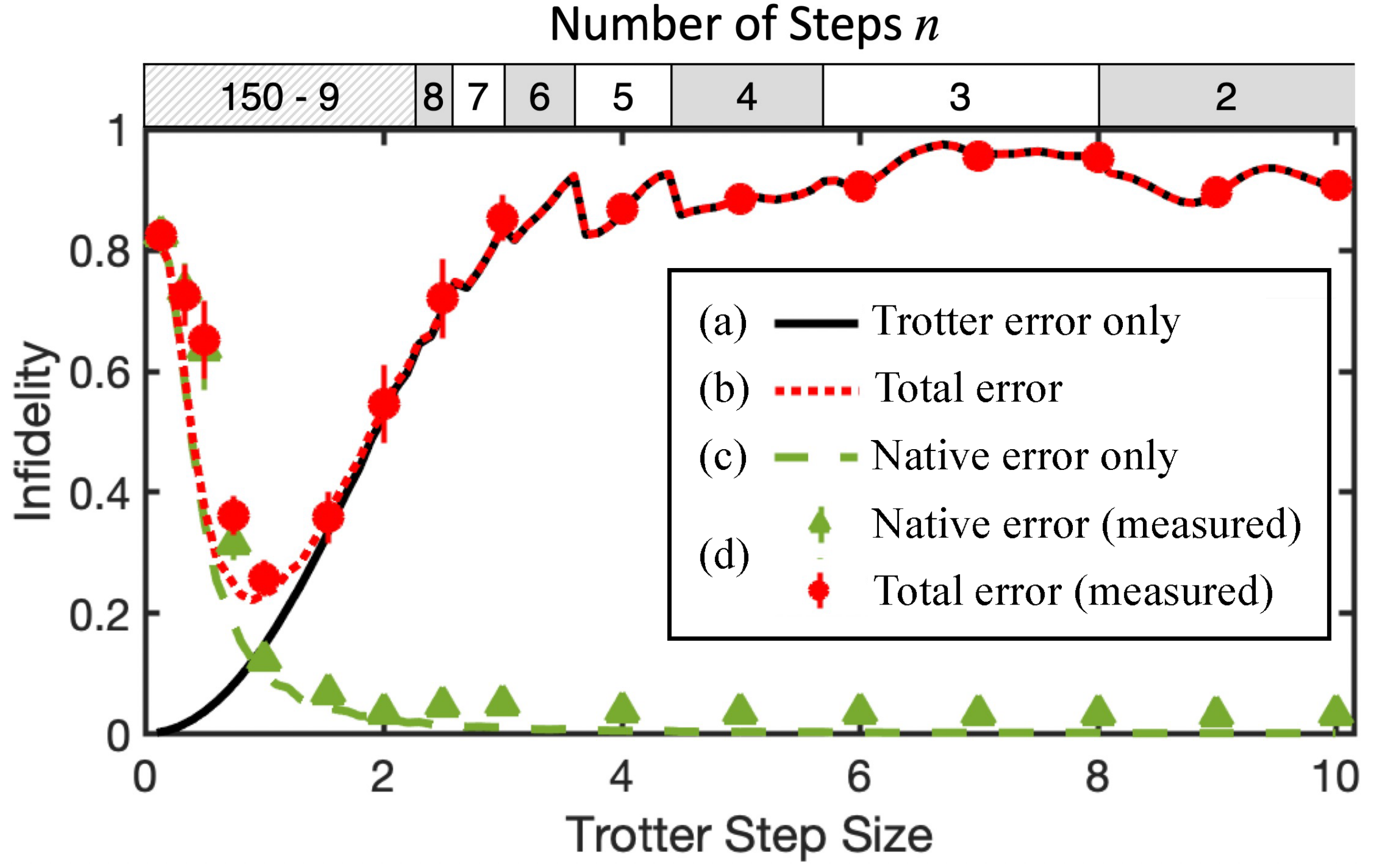}}
\caption{\label{fig:Fig1} (color online) Average simulation infidelity for $12$ spin-coherent states, evolved to $T = 20$ with a range of Trotter step sizes. For a given $\tau$ the unitary time step is iterated $n=T/\tau$ times. (a) Predicted infidelity when comparing perfect Trotterized evolution, $U_{\rm Trot}(\tau)^n$, to perfect LMG evolution, $U_{\rm LMG}(\tau)^n$. (Trotter error only, black solid line). (b) Predicted infidelity when comparing Trotterized evolution with added native errors to perfect LMG evolution. (Total error, red dotted line.)  (c) Predicted infidelity when comparing Trotterized evolution with added native errors to perfect Trotterized evolution. (Native errors only, green dashed line.) (d) Scatter points (red circles and green triangles) are experimentally measured infidelities, in good agreement with theoretical predictions. Error bars show one standard deviation of the mean across the selected initial states.}
\end{figure}

Theoretical study of Trotter expansions suggest errors grow with step size as $\tau^{p+1}$, where $p$ is the order of the product formula (here $p=1$). In the absence of native errors, this implies that one should choose the smallest possible $\tau$ to minimize error when propagating the simulation to a fixed final time $T$. When native errors are present, they are likely to grow with the number of time steps (because each unitary time step is slightly imperfect), and with the total time for which the processor is running (due to the cumulative effect of environmental noise). Our SHAQ processor has native errors of both kinds, though the former are dominant \cite{Lysne2020, Poggi2020}. However, Optimal Control on the SHAQ processor is unbiased in regards to the target map: it finds controls of fixed duration and almost identical fidelity for any unitary target. That means we can use the protocol outlined in \cite{Lysne2020} to generate $U_{\rm LMG}(\tau)$ as well as $U_{\rm Trot}(\tau)$, without explicitly separating the latter into two steps. This effectively simulates the Trotter error for any $\tau$, without introducing additional native errors. Also, the Hilbert space dimension of the SHAQ processor is small enough that we can model its behavior on a classical computer, yet large enough to display non-trivial dynamics such as quantum chaos. 

The trade-off between native and Trotter errors in our experiment is illustrated in Fig. 1. Here, we show the average simulation infidelity for $12$ initially spin-coherent states that evenly cover the spherical phase space of a spin $J = 15/2$. Any infidelity $\leq 1$ is the result of either native errors on their own, Trotter errors on their own, or a combination of both. Also shown are predictions for the average infidelity based on numerical simulations of these scenarios, with native errors modeled using the error model developed in \cite{Poggi2020}. As can be seen, experimental data and theoretical predictions are in close agreement, and both show a clear minimum in infidelity at a distinct Trotter step size, in this case near $\tau=1$. Intuitively, this happens because a small step size requires many steps, leading to large native error. Increasing the step size reduces the number of steps and thus the native error. Eventually a point is reached where the combined effect of native and Trotter errors is minimized, and as the step size is further increased Trotter errors become large and dominant.  

As demonstrated above, native errors set a lower bound on Trotter step size below which simulations become unreliable. Similarly, Trotter errors set an upper bound on the useful step size, with the eventual onset of chaos indicating a definite breakdown in the ability to perform accurate simulations. This makes it essential to find signatures of chaos that can be observed on the quantum processor itself.

\begin{figure*}
[t]\resizebox{18cm}{!}
{\includegraphics{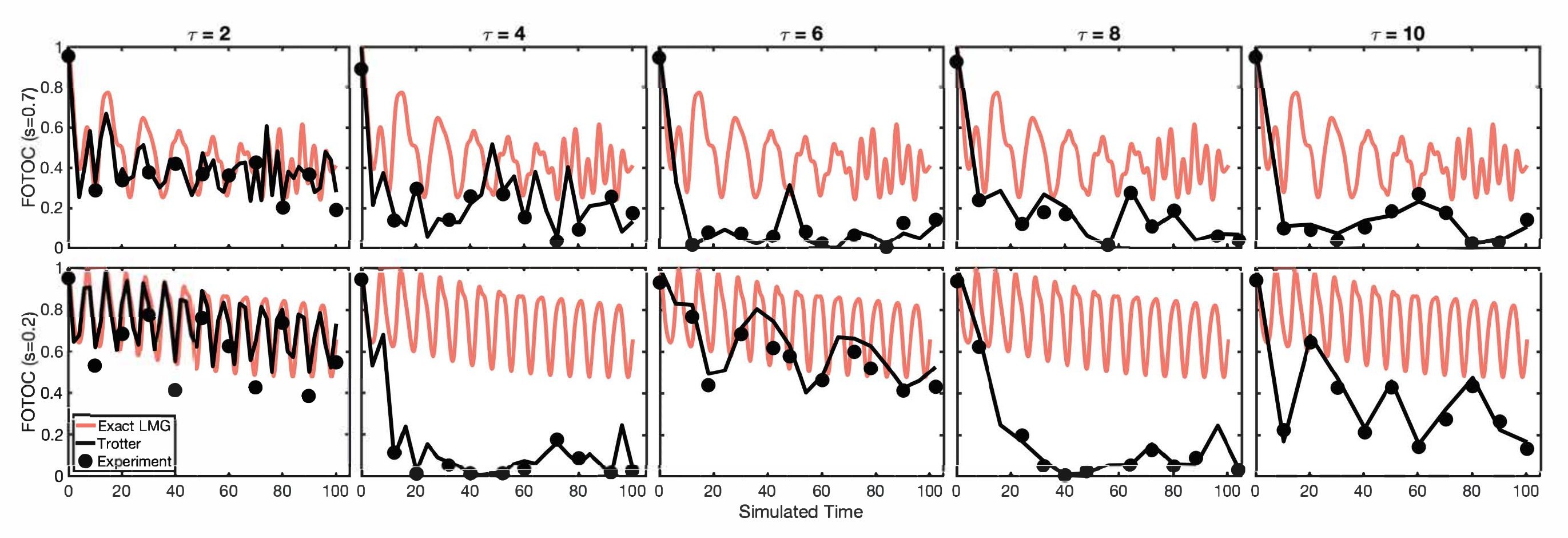}}
\caption{\label{fig:Fig2} (color online)  Measured FOTOC as a function of simulated time for different Trotter step sizes $\tau$. All plots show the FOTOC for a spin-coherent state initially oriented along $\theta=0.833\pi, \varphi=0$, as predicted for the exact LMG model (light red line), and for the exact Trotterized LMG model (solid black line). Scatter points are experimentally measured values of the FOTOC at equal intervals. The top row shows data for $s=0.7$, the bottom row for $s=0.2$}
\end{figure*}

In what remains of this letter, we use an out-of-time-order correlator (OTOC) to determine if chaos is present. OTOCs have been connected to Lyapunov exponents in classically chaotic systems \cite{Maldacena2016, Hashimoto2017}, making it a physically motivated choice when looking for quantum chaos. To measure this OTOC we adopt a procedure detailed in \cite{Blocher2022}. As a first step, we choose a spin-coherent initial state $|\psi_0 \rangle=|\theta,\varphi \rangle=|\hat{n}\rangle$, oriented along the $\hat{n}$  direction in phase space. We then define the OTOC as 
\begin{align}
&F(T)=\langle \psi_0|W^{\dagger}(T) V^{\dagger} W(T) V |\psi_0 \rangle \tag{3},
\end{align}
\noindent where $V=|\psi_0\rangle\langle\psi_0|$ is the projector onto the initial state, $W_0 =e^{-i(J \cdot \hat{n}) \alpha}$ is a rotation about the axis $\hat{n}$, and $W(T) =(U^{\dagger}_{\rm Trot}(\tau))^n W_0 U_{\rm Trot}(\tau))^n $ is the time-evolved version of $W_0$. In the following we choose $\alpha=2\pi/d$ for optimal resolution. The choice of $V$ makes this a fidelity-OTOC, and Eq. (3) simplifies to
\begin{align}
&F(T)=\lvert \langle \psi_0|W(T) |\psi_0 \rangle \rvert^2=\bigg{|}\sum_{n=1}^d P_n (T) e^{i \lambda_n}\bigg{|}^2 \tag{4},
\end{align}
\noindent where $P_n (T)$ are the (measured) populations and $\lambda_n$ are the eigenangles for the eigenstates of $W_0$. Thus, a measurement of $F(T)$ can be obtained by forward-evolving $|\psi_0 \rangle$  to time $T$, measuring populations in the eigenbasis of $W_0$, and substituting in Eq. (4).

Examples of raw FOTOC data for a single initial state are shown in Fig. 2, for different values of $s$ and $\tau$. As expected when the step size is small, the numerically calculated FOTOC is very similar for Trotterized and exact LMG evolution.  Experimental data from the SHAQ processor are in good agreement with theory but show the impact of native errors for small $\tau$. As the Trotter step size is increased the native errors are reduced, but the Trotterized dynamics starts deviating significantly from the exact LMG model. Most notably, for $s=0.7$ and $\tau \ge 4$ where chaos is widespread, the FOTOCs decay to a fraction of their initial value and largely stays there. For $s=0.2$ there is little chaos and the behavior of the individual FOTOCs is more complex.

Exponential decay of the FOTOC has been associated with scrambling and chaos, and we do see this behavior in regimes where chaos is present in the dynamics. However, as has been pointed out in a number of studies \cite{Kidd2021, Xu2020, Pilatowsky-Cameo2020, Pappalardi2018, Hummel2019, Hashimoto2020}, rapid decay of the FOTOC can also be attributed to states initially located near an unstable point in the classical phase space, which obscures the connection between the FOTOC and chaos.  However, in the regular regime rapid decay is followed by periodic revivals of the FOTOC in time, which are absent for chaotic dynamics \cite{Pappalardi2018, Rigol2008, Fortes2019} . To capture this behavior across states and time, we measured the FOTOC for an LMG model with $s=0.7$, for $12$ initially spin coherent states, and sampled its value at $10$ equidistant points from $t=0$ to $t=100$. By averaging over initial states and time samples, we obtained the data points shown in Fig. 3a.  Also included for comparison are illustrations showing the structure of the classical phase space, along with numerical calculations of the average FOTOC value predicted for the exact Trotterized LMG model (i. e., the QKT of Eq. 2b), with and without native errors.  Strikingly, as the step size is increased from  $\tau=1$ to $\tau=10$, the average FOTOC decreases as chaos gradually takes over the classical phase space, saturates at a minimum value when chaos becomes nearly global, and then recovers again as the classical phase space becomes more regular.  

The saturation point for the average FOTOC can be compared against two theoretical predictions. First, the black dash-dotted lines in Fig. 3 shows the expected FOTOC value when averaged over a large set of Haar random states, 
\begin{align}
&\overline{\langle F(T) \rangle}_{\rm Haar}=\frac{1}{d^2+d}\left(|\mathrm{Tr}[W]|^2+d\right) \tag{5},
\end{align}

\begin{figure*}
[t]\resizebox{18cm}{!}
{\includegraphics{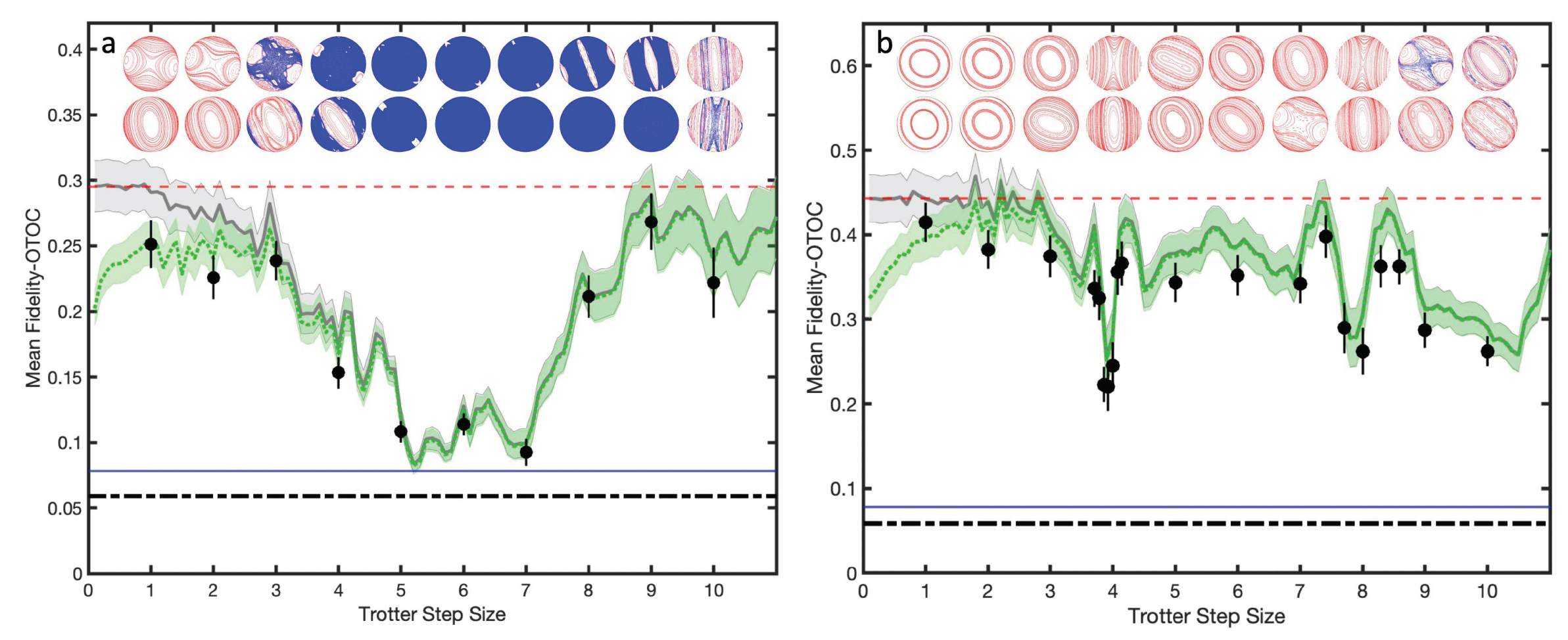}}
\caption{\label{fig:Fig3}(color online) (a) FOTOC values averaged over $12$ initial spin coherent states, sampled at $10$ points during simulation from $t = 0$ to $t =100$, for a range of step sizes. The red dashed line shows the mean FOTOC for the exact LMG model, the solid grey line is a numerical simulation of the exact Trotterized evolution (e, g, a QKT) for $s=0.7$, and the green dotted line is a numerical simulation including native errors. The blue dash-dotted line is the mean FOTOC for states evolved under $10^4$ COE unitaries with $e^{iJ_{\rm z}\pi}$ symmetry. The black dash-dotted line is the Haar average over the FOTOC observable. Black scatter points are experimental measurements of the average FOTOC. Error bars/bands indicate one standard deviation of the mean for all samples. Classical phase space plots at the top show regular (red) and chaotic (blue) regions at the corresponding step size. (b) Average FOTOC values measured or numerically simulated as in (a), but for an LMG model with $s=0.2.$}
\end{figure*}

\noindent This is the value expected if the dynamics were consistent with random evolution. For large-dimensional systems $\overline{\langle F(T) \rangle}_{\rm Haar}$ approaches $0$, but for the relatively small system size of $d=16$  the average FOTOC takes a value of $\overline{\langle F(T) \rangle}_{\rm Haar}=0.0588$ . We see from Fig. 3a that the Trotterized LMG model does not reach this level of randomness even at its most chaotic, saturating instead at $\approx 0.1$.  This discrepancy is accounted for by a combination of finite-size effects, time-reversal, and parity symmetries in the Floquet operator. As seen in Fig. 3a, when we compare the experiment against a numerical average over FOTOCs obtained by randomly choosing evolution operators with the appropriate symmetries, we find better agreement between experimental data and theoretical predictions. See \cite{Supplement} for details.

It should be noted that for the time simulated $(T = 100)$, the mean value of the FOTOC has not reached steady state for the largest step sizes.  In Fig. 3a the mean FOTOC is seen to increase for $\tau>7$  where the classical phase space gradually becomes more regular. In numerical simulations we find that increasing $T$ has minimal effect for step sizes $\tau\leq 7$ , but does in fact decrease (increase) the mean FOTOC for dynamics showing more (less) chaos when $7 \leq \tau \leq 10$. For very long simulations native errors eventually produce a completely mixed state, which for $\alpha=2\pi/d$ yields a FOTOC value of zero \cite{Supplement}. As indicated by the dashed green line in Fig. 3a, our experiment is far from that limit over the range of $\tau$ displayed. 

To further emphasize the connection between FOTOC values and the presence of chaos, we repeated the above experiment for a nonlinear strength $s=0.2$. The results are shown in Fig. 3b. First, we note that in this case chaos is never widespread in phase space for the Trotter steps considered, and correspondingly there is no regime where the mean FOTOC saturates anywhere near the value observed for random evolution. However, the classical phase space does undergo bifurcations near $\tau=4$ and $\tau=8$. These unstable points were studied in \cite{Chinni2022}, and lead to behavior that can be understood by considering how the linear and non-linear rotations change with step size. As $\tau$  is increased, the linear rotation angle changes in a cyclic manner while remaining bounded by ${\rm mod}[(1-s)\tau,2\pi]$. Because of the $e^{iJ_{\rm z}\pi}$ symmetry, step sizes for which $(1-s)\tau\approx n\pi$ ($n$ integer) will reduce the strength of the linear rotation relative to the non-linear rotation, and this situation leads to localized chaos. The behavior is evident for step sizes near $\tau=4$ and $\tau=8$ in Fig. 3b, where the average FOTOC does decrease, but the narrow features are qualitatively different from the deep and broad minimum caused by nearly global chaos in Fig. 3a. Note that there is a similar instability near $\tau=10$ in Fig. 3a, which in this case makes the phase space less chaotic and causes the average FOTOC value to increase.

In summary, our experiments highlight some of the challenges facing quantum simulation on non-error corrected quantum processors. In particular, the interplay between native and Trotter errors can be studied and carefully considered when optimizing performance and minimizing errors. The possibility of chaotic evolution driven by Trotterization has been studied elsewhere \cite{Sieberer2019}, and in our work we have observed it directly in experiments running on a physical quantum processor.  Our work also shows that a well-chosen out-of-time-ordered correlator, the fidelity-OTOC (FOTOC), can be measured and used as a reliable indicator of chaos. Our study has focused on Trotterization, which can change system dynamics from regular and integrable to that of a driven system prone to chaos. Chaos is but one interesting feature of driven systems; outcomes such as stable subharmonic oscillation may also occur for systems exhibiting a time-crystal phase \cite{MunozArias2022}.

Finally, we note that our experiment relies heavily on the ability to model the SHAQ processor on a classical computer. In fact, this is essential to its operation when programming it through Optimal Control, and to the way we measure control fidelity and FOTOC values. Such modeling is already infeasible on many of the NISQ processors in use today. However, Trotter errors can be estimated to some degree through mathematical analysis \cite{Childs2021}, while native processor errors can be estimated through methods such as elided circuits \cite{Arute2019} or Loschmidt echoes \cite{Lysne2020,Peres1984}. As for FOTOC values, these can in principle be accessed on a NISQ processor by evolving forward in time with the map $U_{\rm Trott}(\tau)^n$, applying the map $W_0$, evolving backward with the map $U_{\rm Trott}^{\dagger}(\tau)^n$, and measuring the probability of recovering the initial state. If these steps can be demonstrated in the laboratory, it seems likely that the FOTOC will prove a useful indicator of chaos on a broad range of quantum simulators.

\begin{acknowledgments}
We thank Philip Blocher, Manuel Mu\~{n}oz-Arias, and Ivan H. Deutsch for helpful discussions. This work was supported by the US National Science Foundation Grants PHY -1212308, PHY-1212445, PHY-1307520.
\end{acknowledgments}



\widetext
\clearpage
\setcounter{section}{0}
\setcounter{equation}{0}
\setcounter{figure}{0}
\setcounter{table}{0}
\makeatletter
\makeatother

%
%
%
%
%
%

\begin{center}
\textbf{\large Supplemental Material: Native, Trotter, and chaotic errors in quantum control} \\
\vspace{4mm}
Kevin W. Kuper,$^{1,*}$ Jon P. Pajaud,$^{1}$ Karthik Chinni,$^{2}$ Pablo M. Poggi,$^{2}$ and Poul S. Jessen$^{1}$ \vspace{1mm} \\
 \begin{small}
 $^{1}$\textit{Wyant College of Optical Sciences, University of Arizona, Tucson, Arizona 85721, USA} \\
	 $^{2}$\textit{Center for Quantum Information and Control (CQuIC), Department of Physics and Astronomy, \\
	  University of New Mexico, Albuquerque, New Mexico 87131, USA}\\
	  (Dated: December 19, 2022)
\end{small}\\
\vspace{6mm}
\end{center}

\pagenumbering{gobble}

\twocolumngrid

\section{I.  Haar-Averaged FOTOC}

\noindent We consider the fidelity-OTOC

\begin{equation}
    F = \bra{\psi} W^\dagger V^\dagger W V \ket{\psi}
\end{equation}

\noindent where $V$ is the projector onto the initial state $V=\ketbra{\psi}{\psi}$, and $W$ is a unitary transformation which initially commutes with $V$. In this case, $F$ reduces to

\begin{equation}
    F = |\bra{\psi} W \ket{\psi} |^2.
\end{equation}

\noindent In the following, we will be interested in finding the average $F$ over Haar-random states. To do this, we present a derivation of the more general problem of finding the Haar-average over the quantity $\langle A \rangle \langle B \rangle$, which when $A=W$ and $B=W^\dagger$ becomes our FOTOC $F$. Consider a set of basis states $\{\ket{n}\}$, where $n=1,\dots,d$ with $d$ being the dimension of the system. We can express a random state $\ket{\psi}$ as

\begin{equation}
    \ket{\psi} = U \ket{n}
\end{equation}

\noindent where $U$ is a random unitary operator drawn from the uniform Haar measure, and $n$ is arbitrary. Since the operators $A$ and $B$ may be expressed as

\begin{equation}
    A = \sum_{ij} A_{ij}\ketbra{i}{j}
\end{equation}

\begin{equation}
    B = \sum_{kl} B_{kl}\ketbra{k}{l}
\end{equation}

\noindent we see that
\begin{align}
    \langle A \rangle \langle B \rangle &= \bra{\psi}A\ketbra{\psi}{\psi}B\ket{\psi}\\
    &=\sum_{ijkl}A_{ij}B_{kl}(U^\dagger)_{ni}U_{jn}(U^\dagger)_{nk}U_{ln}\\
    &=\sum_{ijkl}A_{ij}B_{kl}U_{jn}U_{ln}U_{in}^*U_{kn}^*.
    \label{eq8}
\end{align}


\noindent Therefore, we need to compute the Haar-average of the expression $U_{jn}U_{ln}U_{in}^*U_{kn}^*$, which is a fourth-order `monomial' in the matrix elements of $U$. For low-order monomials like these, closed-form expressions are known \cite{suppCollins2006}. 

\vspace{0.8mm}
\noindent \rule{0.5in}{1pt} \vspace{1pt} \\
\noindent {\small $^{*}$ kwgkuper@email.arizona.edu} \\

\noindent In this case, we obtain

\begin{equation}
    \mathbb{E} [U_{jn}U_{ln}U_{in}^*U_{kn}^*] = \frac{1}{d(d+1)}(\delta_{ij}\delta_{lk} + \delta_{jk}\delta_{li}).
\end{equation}

\noindent Thus, we can take the average of Eq. \ref{eq8}, obtaining

\begin{align}
    \overline{ \langle A \rangle \langle B \rangle} &= \frac{1}{d(d+1)} \left(\sum_{il}A_{ii}B_{ll} + \sum_{jl}A_{lj}B_{jl} \right) \\
    &=\frac{1}{d(d+1)}(\Tr(A)\Tr(B)+\Tr(AB))
\end{align}

\noindent with which we may obtain our final result by setting $A=W$ and $B=W^\dagger$

\begin{equation}
    \overline{F}_{\mathrm{Haar}}= \frac{1}{d(d+1)}(|\Tr(W)|^2 + d).
    \label{eq12}
\end{equation}

\section{II.  Choice of Rotation Angle}

\noindent In calculating the FOTOC, we must choose a particular unitary $W$ which commutes with the projector onto the initial state $V=\ketbra{\psi_0}{\psi_0}$. In our case, we have chosen initial states which are spin-coherent, and whose spin is oriented in the direction given by the azimuthal and polar angles $(\theta,\phi)$. Thus, a natural choice for the unitary $W$ is a rotation about an axis similarly oriented by $(\theta,\phi)$:

\begin{equation}
    W = e^{-i(\vec{J}\cdot \hat{n}(\theta,\phi))\alpha}
\end{equation}

\noindent where the rotation $\alpha$ may be arbitrarily chosen. Choosing $\alpha=0$ sets $W$ to the identity, giving us no resolution on the FOTOC in Eq. \ref{eq12}.  To maximize the resolution of $F$, we must choose a rotation angle which minimizes $|\Tr(W)|^2$. This is accomplished by the choice

\begin{equation}
    \alpha = \frac{2\pi}{2J + 1} = \frac{2\pi}{d}
\end{equation}

\noindent which leaves $W$ with pairs of eigenvalues on opposite sides of the complex unit circle (see Fig \ref{fig:Fig1}) and thus $\Tr(W)=0$. The resulting Haar-average FOTOC is then minimized and takes the value 

\begin{equation}
    \overline{F}_{\mathrm{Haar}}=\frac{1}{d+1}.
\end{equation}

\noindent For $d=16$, we have $\overline{F}_{\mathrm{Haar}}\simeq 0.0588$, which is plotted as a black dash-dotted line in Fig. 3 in the main text. 

\begin{figure}
    \centering
    \includegraphics[width=1\linewidth]{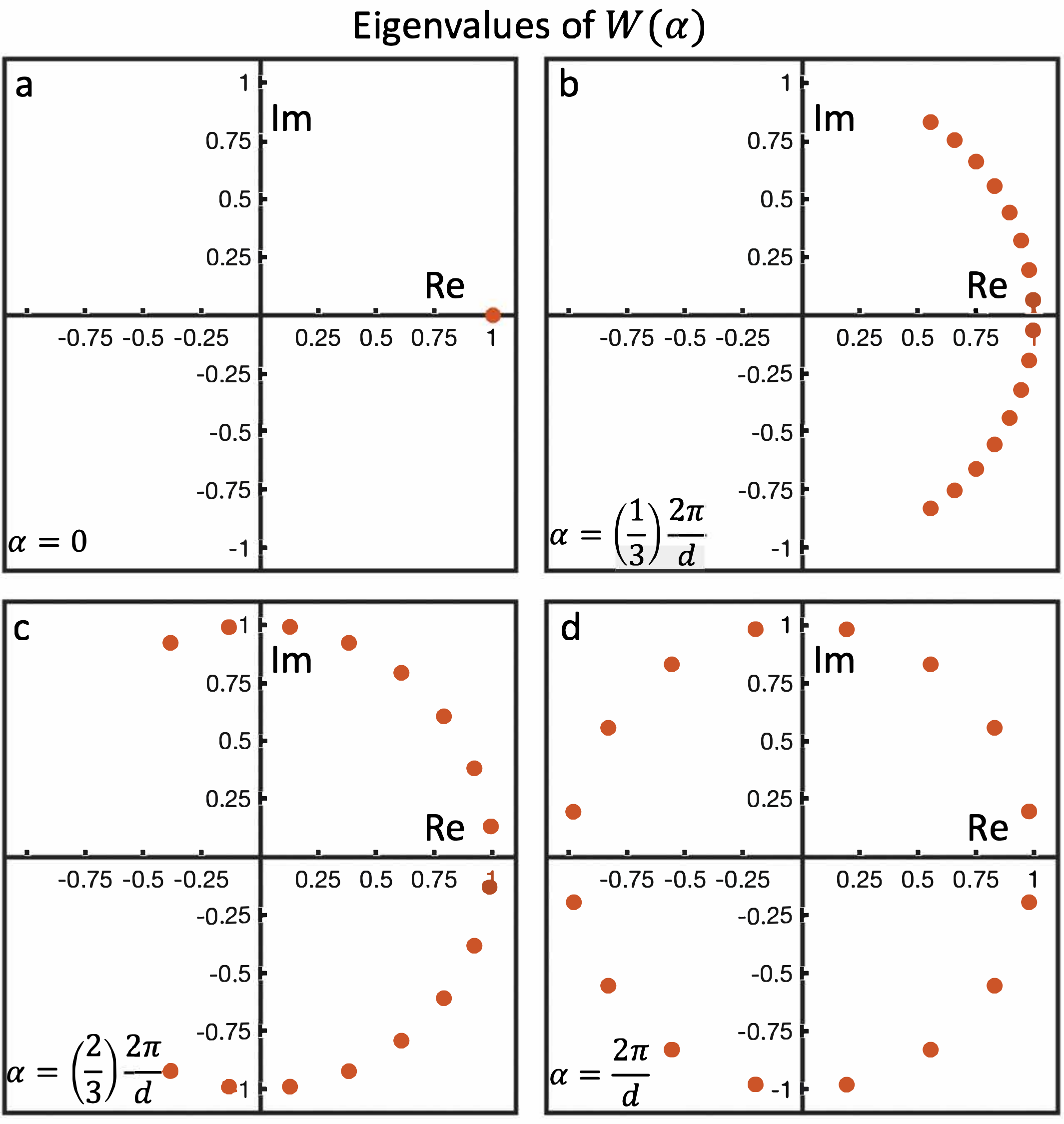}
    \caption{Eigenvalues of $W(\alpha)$ (with $d=16$) plotted on the complex unit circle for values of a) $\alpha=0$, b) $\alpha=\left(\frac{1}{3}\right) \frac{2\pi}{d}$, c) $\alpha=\left(\frac{2}{3}\right) \frac{2\pi}{d}$, and d) $\alpha=\frac{2\pi}{d}$.}
    \label{fig:Fig1}
\end{figure}

\pagenumbering{arabic}
\setcounter{page}{2}

\section{III.  Role of symmetries}

\noindent In the main manuscript we have studied the fidelity-OTOC as an indicator for the onset of chaos in the dynamics of the Trotterized LMG. One of the main features of quantum chaotic dynamics is that it can dynamically generate pseudorandom states in Hilbert space starting from any generic initial configuration. In collective spin systems such as the one studied here, this implies in particular that localized product states (i.e. spin coherent states) become entangled and highly delocalized in phase space. Absent any other constraints, the resulting state can be considered to be taken from the uniform distribution of states in Hilbert space, leading to the calculation presented in the previous section.
\indent However, the specific physical models studied in this work display particular symmetries that constrain the generation of randomness in the dynamics. In particular, the LMG Hamiltonian (cf. Eq. 1 in main text) and its corresponding unitary evolution operator, and that of its Trotterized version, display both time-reversal and parity symmetry for generic choices of the system's parameters.

\subsection{A.  Time reversal symmetry}

\noindent Time-reversal symmetry is one of the basic symmetry groups considered in the context of quantum chaos and its characterization in terms of random matrices \cite{suppHaake1991}. Time-reversal symmetry of a Floquet operator $U$ can be associated with the existence of an antiunitary operator $T$ such that $T^2 = 1$ and $TUT^{-1}=U^\dagger$. This implies that one can always find a basis in which $U$ is symmetric (see Ref. \cite{suppHaake1991}, Sect. 2.12),  which in turn constrains the the values of its matrix elements. Random unitary matrices which are symmetric but otherwise uniformly distributed constitute the Circular Orthogonal Ensemble (COE). For the case of the Kicked Top unitary considered in the main text, the analysis of the time-reversal symmetry was originally presented by Haake \cite{suppHaake1987}.

Notice that the averages we computed analytically  in the preceeding section were performed over the (unconstrained) uniform Haar measure, which is equivalent to considering the Circular \textit{Unitary} Ensemble (CUE). Even though explicit calculations of averages over COE have been studied in the past \cite{suppMatsumoto2012}, the procedure is much more cumbersome even for second order moments, and so we will tackle the problem mostly numerically.


\subsection{B. Parity symmetry} 

\noindent The Floquet operator corresponding to the quantum Kicked Top, that represents the Trotterized LMG evolution, has the form

\begin{equation}
    U_{QKT} = e^{ip J_z}e^{ik J_x^2}
\end{equation}

\noindent Introducing the parity operator $R_z = e^{-i \pi \left(J_z-J\right)}$, we see that

\begin{equation}
    \left[ U_{QKT}, R_z\right] = 0\ \forall\ p,k
\end{equation}

\noindent It is easy to see that the parity operator has only two possible eigenvalues $\pm 1$. As a result of this symmetry, the eigenvectors of $U_{QKT}$ are divided into those which are even ($+1$) under parity and those which are odd ($-1$). Any evolution generated by $U_{QKT}$ will preserve parity, and thus the generation of randomness from any given state will be hindered by this symmetry.

\subsection{C.  Average FOTOC with symmetries and finite size effects} 
\begin{figure}
    \centering
    \includegraphics[width=1\linewidth]{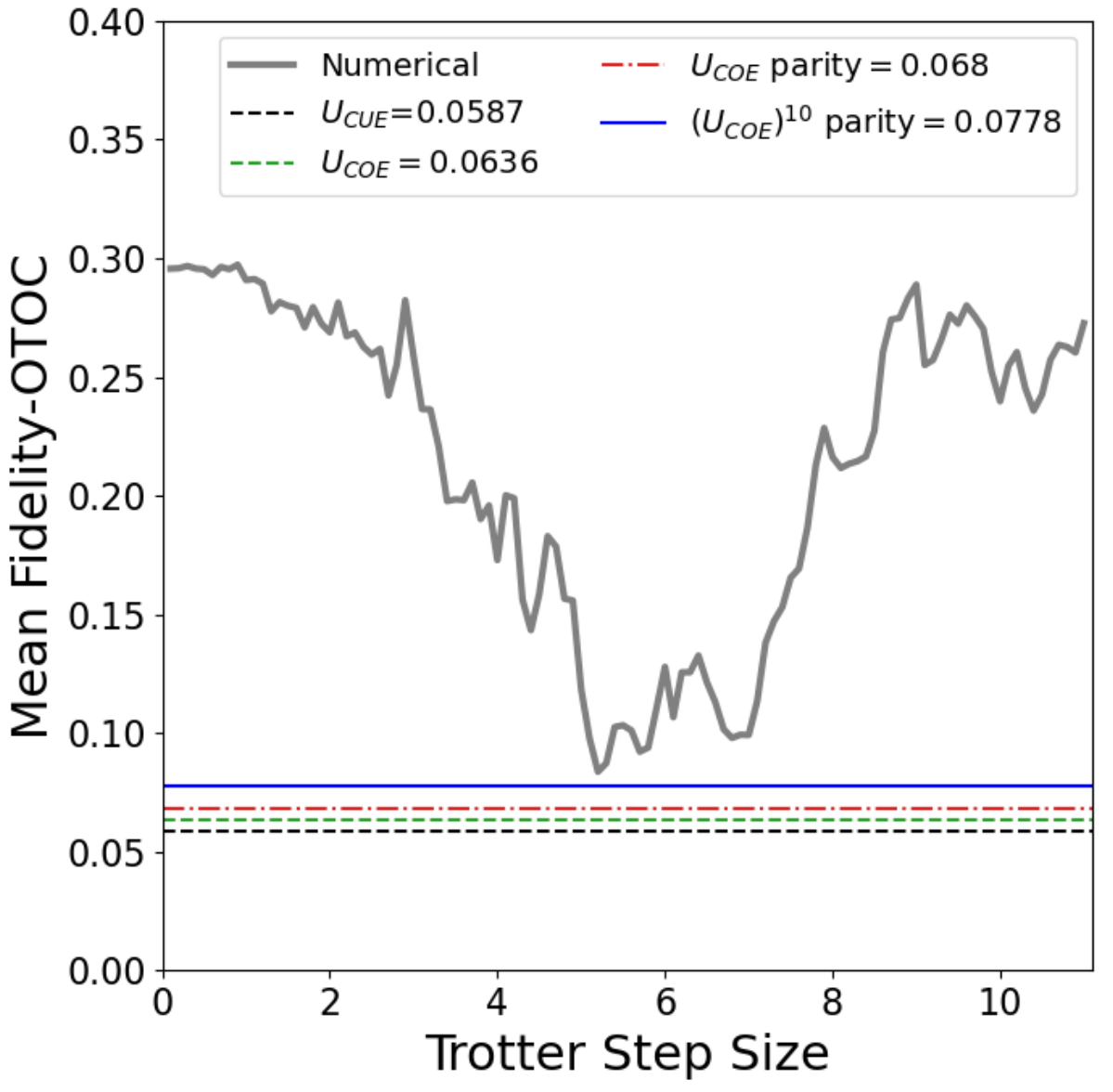}
    \caption{The mean fidelity-OTOC obtained from the Trotterized evolution of the LMG Hamiltonian from $t=0$ to $t=100$ for $s=0.7$ and various values of the Trotter step sizes, identical to the grey-colored curve shown in Fig. 3a of the main text, is compared with predictions of the various random matrix theory ensembles. It can be seen that the additional symmetry constraints increase the value of the mean fidelity-OTOC ($U_{COE}$ vs. $U_{COE}$ vs. $U_{COE}$ parity). In addition, there are finite-size effects, which are taken into account in obtaining the thick blue line. For more details on these finite-size effects, see text and refer to Fig. 3. Note that the thick blue agrees well with the mean fidelity-OTOC obtained from the Trotterized evolution (grey curve) in the chaotic regime.}
    \label{fig:Fig3}
\end{figure}

\noindent We now turn to analyze in closer detail how symmetries affect the expected value of the fidelity-OTOC in the case of chaotic motion. As a rule of thumb, the presence of the symmetries will tend to increase the value of the average FOTOC above the Haar-random case $\simeq 1/(d+1)$, which can be seen as the minimum value. The effect of symmetries can be exaggerated if we make specific choices of $W$ and $\ket{\psi}$. For instance, consider a choice of $W$ such that $W=W_+ \oplus W_-$, and take only states of definite parity, say $\ket{\psi}=\ket{\psi_+}$, which are taken to be uniformly distributed in that subspace. An analogous calculation to that of Section I yields

\begin{equation}
    \overline{F}_{+}= \frac{1}{d_+(d_+ +1)}\left(|\Tr(W_+)|^2 + d_+\right) \xrightarrow[\mathrm{minimum}]{} \frac{1}{d_+ + 1}
\end{equation}

\noindent Naturally, considering this scenario in presence of symmetry is equivalent to performing a calculation on a reduced subspace of definite parity, whose dimension is $d_+ = d/2$. The average FOTOC then is larger than in the unconstrained case. However, in the protocol described in the main text we deal with random choices of $W$, which will tend to wash out this effect to some degree. \\
\indent We now discuss the procedure to obtain the average FOTOC numerically using random matrix theory techniques taking into account the effect of all the symmetries. We first sample $d/2$-dimensional matrices from the COE ensemble and arrange them into block diagonal form $U_{\text{COE},\Pi}(d)\equiv U_{\text{COE}}(d/2) \oplus U_{\text{COE}}(d/2)$ so as to mimic random matrices generated under the parity symmetry constraint. The method to sample a matrix $U$ from the COE is to  first sample a matrix $V$ from CUE using standard techniques and then constructing $U=V^T V$. We then take the average of the FOTOC obtained over the same initial conditions used in obtaining the green/gray curve shown on Fig. 3 in the main text. Finally, we repeat the above process and obtain an average over different random matrices sampled from the COE ensemble with parity symmetry. This can be quantitatively expressed in terms of the following quantity when $U_{\lambda}=U_{COE,\Pi}$
\begin{align}
\label{eq:ratio}
\begin{split}
       \overline{F}_{avg}^{(\lambda)}&\equiv\frac{1}{k}\sum_{j=1}^{k}\mathbb{E}\bigl(\big|\bra{\psi_{0}^{(j)}}U^{\dagger}_{\lambda} W_j U_{\lambda}\ket{\psi_{0}^{(j)}}\big|^2\bigr)\\
    &=\frac{1}{k}\sum_{j=1}^{k}\overline{F}\bigl(|\psi_{0}^{(j)}\rangle \bigr) 
\end{split}
\end{align}
where $\mathbb{E}$ represents the expectation value over the COE matrices and the summation over $j$ is performed to obtain the average over $k$ initial states. The above quantity is shown by the red dash-dotted line shown in Fig. \ref{fig:Fig3}. As can be seen from the figure, this value does not match with the FOTOC of $W$ evolved under the Trotterized LMG in the chaotic regime. We understand this to be a consequence of the fact that we are dealing with a small Hilbert space of dimension $d=16$. To understand this further, we define a quantity $r_{n}$ as shown below
\begin{align}
\label{eq:ratio}
    r_{n}&=\frac{\overline{F}_{avg}^{(\lambda)}}{\frac{1}{k}\sum_{j=1}^{k}\frac{1}{n}\sum_{m=1}^{n}\mathbb{E}(\langle \psi^{(j)}_{0}|(U^{\dagger}_{\lambda})^{m} W_j (U_{\lambda})^{m}|\psi^{(j)}_{0}\rangle\big|^2)}
\end{align}
The above quantity is the ratio of the FOTOC obtained under a single application of the random matrix $U_{\lambda}$ on the initial state averaged over initial conditions ($\overline{F}_{avg}^{(\lambda)}$) to the averaged FOTOC obtained from the application of $n$ number of random unitary ($U_{\lambda}$) matrices on the initial state that is also averaged over the initial conditions. In Fig. \ref{fig:Fig2}, we plot $r_{10}$ as a function of the Hilbert space dimension. As can be noted, for a larger dimensional system, the ratio is $\sim 1$ indicating no difference between averaged FOTOC obtained from a single application of the random matrix $U_{\lambda}$ to the various applications ($(U_{\lambda})^{10}$ to be specific) of these matrices on the initial state. However, for a smaller dimensional Hilbert space such as the one analyzed in this manuscript ($d=16$), the ratio is smaller than one, and we find that the quantity in the denominator of $r_{10}$ matches better with the theoretically/experimentally obtained curve. Note that the denominator in $r_{10}$ is the quantity of interest because we also take $10$ sample points along the evolution for $T=100$ for various of Trotter step sizes to obtain the green/grey colored curves in Fig. 3 of the main text. We find that this averaged FOTOC over $10$ applications of COE matrices with parity constraint ($U_{COE,\Pi}$) agrees with the grey/green colored curve for the chaotic values of the Trotter step size as shown by the solid blue line in Fig. \ref{fig:Fig3}.\\

\begin{figure}
    \centering
    \includegraphics[width=1\linewidth]{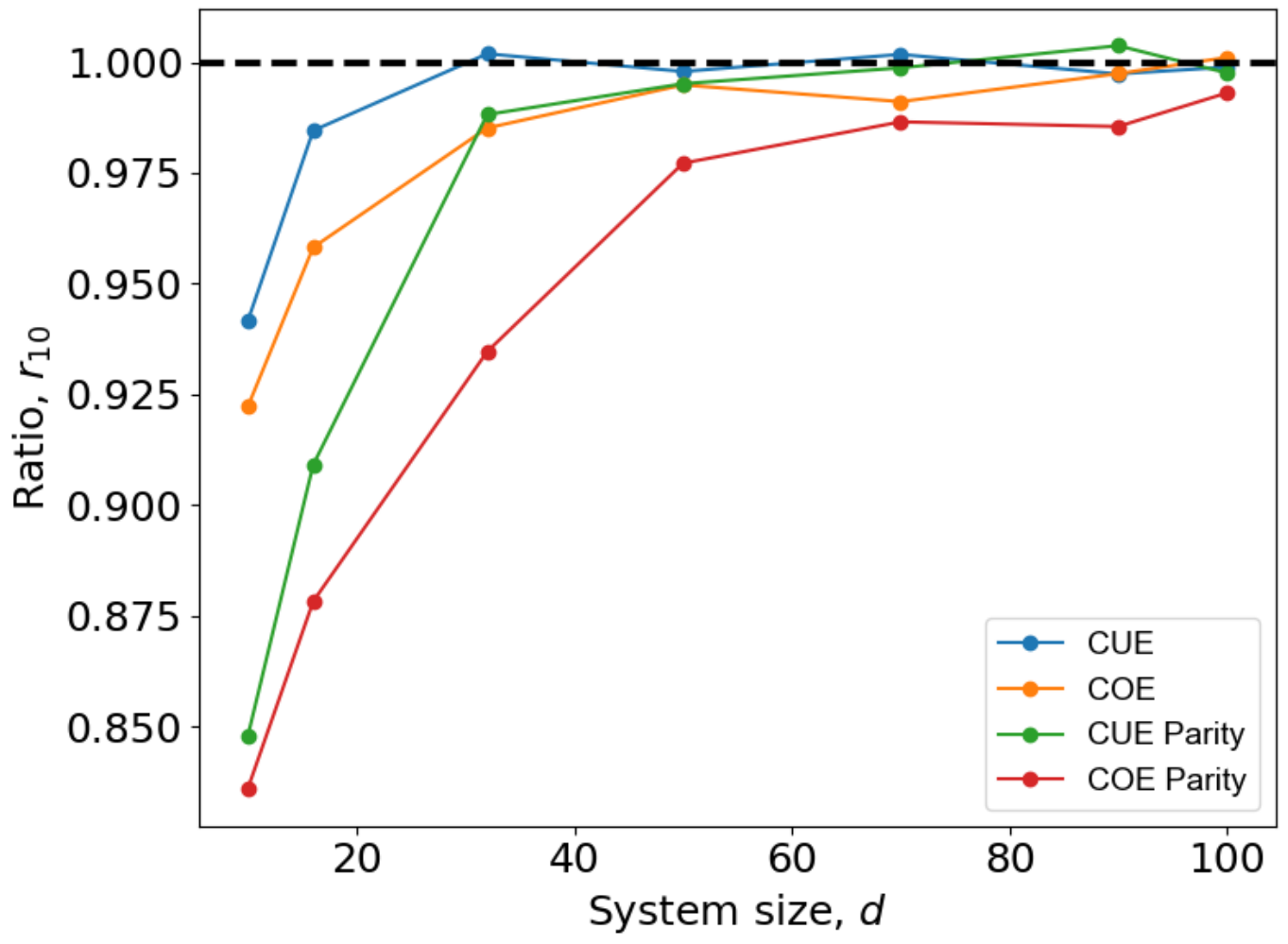}
    \caption{The ratio $r_{10}$(see Eq. \ref{eq:ratio}) of the FOTOC obtained from the application of a single unitary to the averaged FOTOC obtained from the application of 10 random unitaries sampled from the appropriate ensemble has been plotted above. As it can be seen, the ratio approaches $1$ as the system size is increased.}
    \label{fig:Fig2}
\end{figure}


\begin{thebibliography}{3}

\bibitem{Arute2019}
{F. Arute, K. Arya, R. Babbush, D. Bacon, J.C. Bardin, R. Barends, R. Biswas, (...) , and J. M. Martinis, Quantum Supremacy using a programmable superconducting processor, Nature, {\bf 574}, 505 (2021).}

\bibitem{Wu2021}
{Y. Wu, W.-S. Bao, S. Cao, F. Chen, M.-C, Chen, X. Chen, T.-H Chung, H. Deng, Y. Du, D. Fan, M. Gong, J.-W. Pan, \textit{et. al.}, Strong quantum computational advantage using a superconducting quantum processor, Phys. Rev. Lett. {\bf127}, 180501 (2021).}

\bibitem{Feng2022}
{Feng Pan, Keyang Chen, and Pan Zhang, Solving the Sampling Problem of the Sycamore Quantum Circuits, Phys. Rev. Lett. {\bf 129}, 090502 (2022).}

\bibitem{Preskill2018}
{John Preskill, Quantum Computing in the NISQ era and beyond, Quantum {\bf 2}, 79 (2018).}

\bibitem{Baldwin2022}
{C. H. Baldwin,K. Mayer, N. C. Brown, C. Ryan-Anderson, and David Hayes, Re-examining the quantum volume test: Ideal distributions, compiler optimizations, confidence intervals, and scalable resource estimations, Quantum {\bf 6}, 707 (2022).}

\bibitem{Lloyd1996}
{Seth Lloyd, Universal Quantum Simulators, Science {\bf 273}, 1073 (2021).}

\bibitem{Childs2021}
{Andrew M. Childs, Yuan Su, Minh C. Tran, Nathan Wiebe, and Shuchen Zhu, Theory of Trotter Error with Commutator Scaling, Phys. Rev. X {\bf 11}, 011020 (2021).}

\bibitem{Heyl2019}
{Markus Heyl, Philipp Hauke, and Peter Zoller, Quantum localization bounds Trotter Errors in digital quantum simulation, Sci. Adv. 2019 {\bf 5}:eaau8342 (2019)}

\bibitem{Sieberer2019}
{L. M. Sieberer, T. Olsacher, A. Elben, M. Heyl, P. Hauke, F. Haake, and P. Zoller, Digital Quantum Simulation, Trotter Errors, and quantum chaos of the kicked top, npj Quantum Information {\bf 5}, 78 (2019).}

\bibitem{Kargi2021}
{C. Kargi, J. P., Dehollain, F. Henriques, L. M. Sieberer, T. Olsacher, P. Hauke, M. Heyl, P. Zoller, and N. K. Langford, Quantum Chaos and Universal Trotterization Behaviors in Digital Quantum Simulations, https://arxiv.org/abs/2110.11113 (2021}

\bibitem{Smith2013}
{A. Smith, B. E. Anderson, H. Sosa-Martinez, C. A. Riofrio, I. H. Deusch, and P. S. Jessen, Phys. Rev. Lett. {\bf 111}, 170502 (2013).}

\bibitem{Anderson2015}
{B. E. Anderson, H. Sosa-Martinez, C. A. Riofrio, I. H. Deusch, and P. S. Jessen, Phys. Rev. Lett. {\bf 114}, 240401 (2015).}

\bibitem{Lysne2020}
{N. K. Lysne, K. W. Kuper, P. M. Poggi, I. H. Deutsch, P. S. Jessen, Small, Highly Accurate Quantum Processor for Intermediate Depth Quantum Simulations, Phys. Rev. Lett., {\bf 124} 230501 (2020).}

\bibitem{Lipkin1965}
{H. J. Lipkin, N. Meshkov, and A. J. Glick, Nucl. Phys. {\bf 62} 188 (1965).}

\bibitem{Zibold2010}
{T. Zibold, E. Nicklas, C. Gross, and M. K. Oberthaler, Phys. Rev. Lett., {\bf 105} 204101 (2010).}

\bibitem{Haake1987}
{F. Haake, M. Kus, and R. Scarf, Classical and quantum chaos for a kicked top, Zeitschrift f\"{u}r Physik B Condensed Matter, {\bf 65} 381-395 (1987).}

\bibitem{Chaudhury2009}
{S. Chaudhury, A. Smith, B. E. Anderson, S. Ghose, and P. S. Jessen, Quantum Signatures of Chaos in a kicked Top, Nature, {\bf 461} 768 (2009).}

\bibitem{Poggi2020}
{P. M. Poggi, N. K. Lysne, K. W. Kuper, I. H. Deutsch, and P. S. Jessen, Quantifying the Sensitivity to Imperfections in Analog Quantum Simulation, PRX Quantum, {\bf 1} 020308 (2020).}

\bibitem{Maldacena2016}
{J. Maldacena, S. Shenker, and D. Stanford, A bound on chaos, Journal of High Energy Physics {\bf 2016}, 106 (2016).}

\bibitem{Hashimoto2017}
{K. Hashimoto, K. Murata, and R. Yoshii, Out-of-Time-order correlators in quantum mechanics, Journal of High Energy Physics {\bf 2011}, 138 (2017).}

\bibitem{Blocher2022}
{P. Blocher, S. Asaad, V. Mourik, M. Johnson, A. Morello, and K. M\o lmer, Measuring out-of-time-ordered correlation functions without reversing time evolution, Phys. Rev. A {\bf 106}, 042429 (2022).}

\bibitem{Kidd2021}
{R. Kidd, A. Safavu-Naini, and J. Coeney, Saddle-point scrambling without thermalization, Phys. Rev. A {\bf 103}, 033304 (2021).}

\bibitem{Xu2020}
{T. Xu, T. Scaffidi, and X. Cao, Does scrambling equal chaos?, Phys. Rev. Lett. {\bf 124}, 140602 (2020).}

\bibitem{Pilatowsky-Cameo2020}
{S. Pilatowsky-Cameo, J Ch\'avez-Carlos, M. Bastarrachea-Magnani, P. Str\'ansk\'y, S. Lerma-Hern\'andez, L. F. Santos, and J. G. Hirsch, Positive quantum lyapunov exponents in experimental systems with a regular classical limit, Phys. Rev. E {\bf 101} 010202 (2020).}

\bibitem{Pappalardi2018}
{S. Pappalardi, A. Russomanno, B. Z\u unkovi\u c, F. Iemmini, A. Silva, and R. Fazio, Scrambling and Entanglement spreading in long Range spin chains, Phys. Rev. B {\bf 98}, 134303(2018).}

\bibitem{Hummel2019}
{Q. Hummel, B. Geiger, J. Urbina, and K. Richter, Reversible quantum information spreading in many-body systems near criticality, Phys. Rev. Lett. {\bf 123}, 160401 (2019).}

\bibitem{Hashimoto2020}
{K. Hashimoto, K. Huh, K. Kimb, and R. Watanabea, Exponential growth of out-of-time-order correlator without chaos: inverted harmonic oscillator, Journal of High Energy Physics {\bf 11}, 068 (2020).}

\bibitem{Rigol2008}
{M. Rigol, V. Dunjko, and M. Olshanii, Thermalization and its mechanism for generic isolated quantum systems, Nature {\bf 452}, 854 (2008).}

\bibitem{Fortes2019}
{E. Fortes, I. Garc\'ia-Mata, R. Jalabert, and D Wisniack, Gauging classical and quantum integrability through out-of-time-ordered correlators, Phys. Rev. E {\bf 100}, 042201 (2019).}

\bibitem{Supplement}
{See the Supplemental Material at http://link.aps.org for additional details, which includes Refs. \cite{Collins2006, Haake1991, Haake1987, Matsumoto2012}.}

\bibitem{Collins2006}
{B. Collins and P. \'Sniady, Integration with respect to the Haar measure on unitary, orthogonal and symplectic group, Communications in Mathematical Physics {\bf 264}, 773 (2006).}

\bibitem{Haake1991}
{F. Haake, Quantum Signatures of Chaos, in Quantum Coherence in Mesoscopic Systems (Springer, 1991, pp.583-595.}


\bibitem{Matsumoto2012}
{S. Matsumoto, General moments of matrix elements from circular orthogonal ensembles, Random Matrices: Theory and Applications {\bf 1}, 1250005 (2012).}

\bibitem{Chinni2022}
{K. Chinni, M. H. Mu\~noz-Arias, I. H. Deutsch, and P. M. Poggi, Trotter errors from dynamical structural instabilities of Floquet maps in quantum simulation, PRX Quantum {\bf 3}, 010351 (2022).}

\bibitem{MunozArias2022}
{M. H. Mu\~noz-Arias, K. Chinni, and P. M. Poggi, Floquet time crystals in driven spin systems with all-to-all p-body interactions, Phys. Rev. Research {\bf 4}, 023018 (2022).}

 \bibitem{Peres1984}
 {A. Peres, Stability of quantum motion in chaotic and regular systems, Phys. Rev. A  {\bf 30}, 1610 (1984).}

\end{thebibliography}

\begin{thebibliography}{11}
\bibitem{suppCollins2006}
{B. Collins and P. \'Sniady, Integration with respect to the Haar measure on unitary, orthogonal and symplectic group, Communications in Mathematical Physics {\bf 264 }, 773 (2006).}

\bibitem{suppHaake1991}
{F. Haake, Quantum Signatures of Chaos, in Quantum Coherence in Mesoscopic Systems (Springer, 1991, pp.583-595.)}

\bibitem{suppHaake1987}
{F. Haake, M. Kus, and R. Scarf, Classical and quantum chaos for a kicked top, Zeitschrift f\"{u}r Physik B Condensed Matter, {\bf 65} 381-395 (1987).}

\bibitem{suppMatsumoto2012}
{S. Matsumoto, General moments of matrix elements from circular orthogonal ensembles, Random Matrices: Theory and Applications {\bf 1}, 1250005 (2012).}

\end{thebibliography}
\end{document}